Molecular Dynamics Simulation of Miscibility in Several Polymer Blends

**Amirhossein Ahmadi and Juan J. Freire***

*Departamento de Ciencias y Técnicas Fisicoquímicas, Facultad de Ciencias,*

*Universidad Nacional de Educación a Distancia, 28040 Madrid, Spain.*

Short Title: Simulation of Miscibility in Blends

ABSTRACT: The miscibility in several polymer blend mixtures (polymethylmethacrylate/polystyrene, (1,4-cis) polyisoprene/polystyrene, and polymethylmethacrylate/polyoxyethylene) has been investigated using Molecular Dynamics simulations for atomistic representations of the polymer chains. The trajectories obtained from simulation boxes representing the mixtures have been analyzed in terms of the collective scattering structure function. The Flory-Huggins parameter is determined from fits of the simulation results for this function to the random phase approximation expression. The numerical values of this parameter and its variation with temperature obtained with this procedure show a general qualitative and quantitative agreement with existing experimental data for the different systems. These results together with those previously obtained for the polyvylmethylether/polystyrene blends with the same method are compared with data yielded by other computational simpler approaches.

**Introduction**

The prediction of miscibility in mixtures is interesting both from the theoretical and technological points of view.[1] Usually, this property is characterized by the Flory-Huggins parameter, $\chi$.[2] This parameter gives a theoretical description of the phase separation curves for polymers of different chain lengths and compositions. Parameter $\chi$ can be experimentally measured from thermodynamic properties.[3] Actually, experimental data of $\chi$ for different systems show variation with composition, and sometimes with chain length, what underline some deficiencies of the Flory-Huggins theory. Even if these problems are considered, a quantitative measurement of parameter $\chi$ for a given system is generally considered the best indication of its miscibility performance from the macroscopic point of view. Experimental values of $\chi$ can be obtained from data of the neutron scattering structure function. Usually, $\chi$ is numerically fitted to give the best reproduction of the structure function to a theoretical expression given by the random phase approximation (RPA).[4]

A similar approach can be used to obtain theoretical predictions of $\chi$ with numerical data extracted from numerical simulations. Atomistic representations of the mixtures can be built in a simulation box, including periodic boundary conditions. Molecular Dynamics (MD) constitutes the most convenient method of simulation for systems composed by molecules of relatively complex chemical structure.[5] These simulations should use appropriate forcefields to describe the interactions between different atoms. The MD simulations yield numerical trajectories of polymer blends. From these coordinates, "collective scattering" functions can be easily evaluated and the $\chi$ parameter can be evaluated by comparison with RPA predictions, following a procedure similar to the method used to analyze the experimental structure functions. This approach has been followed in some recent investigations for polymers with relatively simple molecular chemical structures as polypropelene/polyethylene[6] and polyisobutylene /polypropylene.[7]

In a previous work,[8] we reported results obtained for the structurally more complex polyvylmethylether/polystyrene blend, *PVME/PS*, which has been broadly studied form the experimental point of view, detailing the choice of different simulation options, parameters and macroscopic variables that we have employed in order to perform an efficient prediction of $\chi$. These results were obtained using short chains and a decreased density. They were in general agreement with the experimental data,[3,9] showing compatible mixtures and a lower critical solution temperature.

In the present work, we report calculations for three other different blends: polymethylmethacrylate/polystyrene, *PMMA/PS*, (1,4-cis) polyisoprene/polystyrene, *PI/PS,* and polymethylmethacrylate/polyoxyethylene, *PMMA/POE*, using similar prescriptions. Precise values of the Flory-Huggins parameter $\chi$ for all these mixtures have been documented from neutron scattering data.[3] *PMMA/PS* blends show slightly positive and temperature-decreasing $\chi$ values.[10] Therefore, long molecular weight *PMMA/PS* blends should show immiscibility and an upper critical solution temperature. *PI/PS* systems behave similarly, but they exhibit considerable higher values of $\chi$.[11] Finally, PMMA/POE blends are considered to be miscible[12] and, actually an upper critical solution temperature has been located for relatively high molecular weights.[13] Neutron scattering data of $\chi$ are partially consistent with this description, though the available results correspond to deuterated PMMA chains. These particular values are small and negative and do not show any noticeable variation with temperature.[14]

The ability of our numerical results to reproduce experimental data for the different systems gives an idea of the usefulness of the method to obtain simulation estimations of $\chi$ in comparison with other numerically simpler procedures. In particular, we consider in our discussion the results that we have previously obtained[15] from a direct calculation of interaction energies and coordination numbers between polymer units, proposed time ago.[16] Moreover, we also take into consideration values obtained from our theoretical method based in the evaluation of binary interaction integrals between pairs of chains.[15] Both methods have been shown to be especially sensitive to details in the potential describing the interaction between atoms.

**Numerical Methods**

Full details on the methods to build chains and construct the simulation box have been previously described.[8] We choose atactic configurations for the *PS* and *PMMA* chains. For *PS, PI* and *PMMA*, we build chains of 3 repeat units. For *POE*, the chains include 4 repeat units. The average radii of gyration of all these chains are in the range 3.7±0.2 Å, as it is also the case for the previously investigated *PVME* chains, composed of 5 repeat units.

We have been able to construct relaxed initial configurations of moderate total energy, i.e. without serious overlapping between chains, in simulation boxes of size $L$ with periodic boundary conditions with a given mean density, $\rho$. We include 5 chains of each type in the box. This value is compatible with a reasonable use of our computational resources and may be sufficient to reproduce the realistic interactions between chains in the small region of the space needed to explore interactions represented by the mean-field parameter $\chi$. Using longer chains corresponding to moderately high molecular weight polymers would require setting high values of $L$. Moreover, longer chains have slower dynamics, reducing the mobility of the systems. All these factors would imply a significant increase of the computer time required to achieve equilibrated samples and to obtain sufficiently long production trajectories. Furthermore, short chains allow for the construction of homogeneous systems even in the case of blends with moderately positive values of $\chi$.

MD simulations are carried out using the "compass" forcefield.[17] "Compass" is based in ab initio quantum mechanical calculations, parameterized to be consistent with condensed phase properties and was shown to give the fastest equilibration and best reproduction of experimental data for the *PVME/PS* blends. Partial charges are set according fixed bond increment rules assigned in the forcefield files and also with the "charge equilibration" method.[18] Differences between the charges obtained with these two methods are important for some atoms.[8] The simulations are performed at constant temperatures with the help of the Andersen thermostat,[5] valid for static properties. We use time steps of 1.5 fs. for the *PMMA/PS* and *PI/PS* systems. This time is higher than those usually employed to study dynamical properties, 0.5 or 1 fs., but we have verified that it is able to give stable trajectories. However, in the case of the *PMMA/POE* systems, we have only obtained stable trajectories if the time step is reduced to 1 fs.

It should be noted that experimental data for the tracer diffusion coefficient of PS are smaller than $10^{-11}$cm$^2$s$^{-1}$, or $10^{-4}$Å$^2$ns$^{-1}$, for *PS* samples of moderately small molecular weight, $M_w$=4500 g./mol, at 150 ºC.[19] Consistently to these data, we expect that the mobility of the interacting molecules in the present systems should be very small when we consider realistic melt densities close to $\rho=10^3$kg/m$^3$, even when we only include short chains in the simulation boxes. According with exploratory results for the mean squared displacement of the center of masses of PS chains in a *PVME/PS* reported previously,[8] we have to decrease density to the smaller value of $\rho\cong 0.7\times 10^3$ kg/m$^3$ in order to give enough mobility and to achieve an adequate equilibration of this type of systems using a reasonable amount of computational time. This density value is also adopted in the present work.

We have verified that, with these specifications, both total energy and temperature oscillate around stabilized mean values in the final trajectories used for the calculation of properties. Equilibration times of 4-8 ns. are typically required for most systems. We perform several (4-7) MD runs from different equilibrated samples, In each run, the system coordinates are saved and included in the statistical samples every 2000 steps. Our final statistics combine the results of the different runs, and typically have to cover a total of 48-72 ns. in order to obtain a sufficient accuracy in the final results.

The collective scattering structure function is computed from the simulation trajectories as a configurational average,

$$S(q) = n_s^{-1} < \sum_i^{n_s} \sum_j^{n_s} f_i f_j \exp(i\mathbf{q}.\mathbf{R}_{ij}) > \quad (1)$$

where $n_s$ is the total number of scattering units. The terms $f_k$ are scattering factors, describing the scattering contrast between different types of atoms. Adopting the simplest description, we consider that all the non-hydrogen atoms contained in the *A/B* blend simulation box are identical scatters and we assume that they have opposite sign for atoms in *A* or *B* chains, with $S(q)=0$ for $q=0$. The latter condition is set so that no scattering is observed at a macroscopic wavelength. Therefore,

$$f_i = n_B / n_S \quad (2a)$$

if atom *i* belongs to an A chain, or

$$f_i = -n_A / n_S \quad (2b)$$

if atom *i* belongs to a B chain ($n_A$ and $n_B$ are the total number of scatters in A and B chains). $\mathbf{R}_{ij}$ is the vector joining centers *i* and *j* in a given configuration. Finally, $\mathbf{q}$ is the scattering vector, whose components are conditioned by the box size,

$$q_k = (2\pi/L) n_k, \quad k \equiv x, y, z, \quad n_k = 1, 2, 3.... \quad (3)$$

**Results and Discussion**

In Figures 1-3 we show representations of $S(q)^{-1}$ vs. $q^2$ results obtained from the MD runs with $T$=400 K with forcefield assigned partial charges for the three different types of blends. Also, two other temperatures, 350 and 450 K, have been considered to cover the realistic range of temperatures for which experimental data have been reported.[3] The simulation points always tend to a positive extrapolated ordinate at $q$=0, confirming the systems homogeneity.

We compare the simulation data with the predictions from the RPA equation[4] that, for the present definition of scattering units can be writing as

$$S^{-1}(q) = \frac{1}{N_A \Phi_A P_A(q)} + \frac{1}{N_B \Phi_B P_B(q)} - 2\chi/V_0 \qquad (4)$$

$V_0$ is the volume of each scatter. Parameter $\chi$ corresponds to a given microscopic reference volume, $V_R$. Therefore,

$$V_0 = V_R / \left( L^3 / n_S \right) \qquad (5)$$

Volume fractions are obtained from the total number of scatters of each type, $\Phi_i = n_i / n_S$. Finally, we should also evaluate the form factors of the two types of chains, according to the expression

$$P_i(q) = (N_i)^{-2} < \sum_i^{N_i} \sum_j^{N_i} exp(i\mathbf{q}.\mathbf{R}_{ij}) > \qquad (6)$$

where the average extend over the different $i$ chains in all the box configurations along the simulation trajectories. ($N_i$ is the total number of scatters in any $i$ chain).

In Figures 1-3, we also plot the RPA curves corresponding to the choices $\chi/V_0$ =-0.05, $\chi/V_0$ =0 and $\chi/V_0$ =0.05. It can be observed that the simulation data are always close to these predictions, showing the same general curvature features that are also exhibited by the RPA predictions. These features are similar for a given blend at all temperatures. Thus, the *PMMA/PS* blends show monotonous curves close to a linear behavior. However, the *PI/PS* systems show a strong curvature downwards, more marked than that observed previously for the *PVME/PS* results.[8] Finally, the *PMMA/POE* curves show a moderate curvature at higher values of $q$.

However, some peculiarities of the simulation points in the different blends should be commented. Thus, the *PI-PS* simulation data show some systematic deviations upwards respect to the RPA curves in the range of $q>1$ Å$^{-1}$. The accommodation of chemical groups at short distances seems to yield an apparent decrease of $\chi$ in the range of distances smaller than 6 Å, though the results obtained at smaller $q$ show a consistent prediction of clearly positive $\chi$ values. The same behavior is observed at other temperatures and also when the charge equilibration method is employed to assign partial charges. In the case of the *PMMA/POE* blends, the simulation curves have a marked sigmoidal aspect. (This feature is also suggested by the *PMMA/PS* data, though deviations with respect to the theory are smaller in this case.) It should be also mentioned that the *PMMA/POE* systems have been harder to equilibrate. Moreover, the simulation results for the *PMMA/POE* systems obtained with the charge equilibration method show even more important deviations with respect to the RPA predictions and a higher statistical noise, see Figure 4. Similar conclusions are obtained for this particular blend at other temperatures.

$\chi$ can be calculated from the simulation results, $S_{sim}(q)$, at a given value of $q$ that have been evaluated according to eq. (1)

$$\chi = \left[ S_{RPA}^{-1}(q, \chi = 0) - S_{sim}^{-1}(q) \right] V_0 / 2 \tag{7}$$

where $S_{RPA}(q,\chi=0)$ is the prediction given by the RPA for S(q=0) for $\chi=0$, i.e. with the first two terms on the right hand of eq. (4). We obtain $\chi$ as the arithmetic mean of the values calculated from eq. (7) with the different values of $q$.

In Table 1 we report the numerical values of $\chi$ obtained following this procedure for the different systems at several temperatures, with partial charges assigned by the forcefield or calculated with the charge equilibration method. Differences between both methods to obtain charges are small for the *PMMA/PS* and the *PI-PS* blends. However, they are more important for the *PMMA/POE* systems, where the charge equilibration method values have higher error bars. The differences between results for $\chi$ obtained with the two charge methods do not follow a systematic pattern for any of these three blends. In our previous work for *PVME/PS* chains, however, we obtained moderately negative $\chi$ values with the charge equilibration method and slightly positive $\chi$ values, actually closer to the slightly negative experimental results, with forcefield assigned charges. Incidentally, the arithmetic means obtained with the data contained in these two sets gave results in good agreement with the experiments.[8]

In Figure 5 we have plotted the arithmetic means obtained from the values of $\chi$ calculated with the two methods to assign charges for the different systems and temperatures. As discussed previously, differences between the results calculated with the both methods are not systematic, except for the previously studied *PVME/PS* systems, whose mean values[8] are also included in Figure 5 for the sake of comparison with the other blends. In the presently investigated systems, evaluating a mean value between both methods may be simply considered as a way to improve statistics and, therefore, we consider the results from both methods as independent samples. These mean values are compared with experimental data obtained form neutron scattering experiments for the different blends, summarized in terms of curves fitted for the considered temperature ranges.[3]

It can be observed that, even though the error bars are significant, the simulation results for $\chi$ are relatively close to the experimental values. Moreover, the data suggest a variation with temperature close to the actual experimental behavior of the different systems, except in the case of the less accurate results for the *PMMA/POE* blends. (We should remark that, as pointed out in the Introduction, the experimental temperature behavior of these blends has not been totally clarified.) However, we should summarize some previously commented methodological points that have to be taken in consideration to judge the performance of the simulation results for the *PI/PS* and *PMMA/POE* systems. Thus, we should have in mind that the *PI/PS* data are calculated from means of the results calculated form eq. (7) after subtracting the 3 highest-$q$ points (short distance range points), for which systematic upward deviations are obtained. Otherwise, we obtain very small absolute values of $\chi$ (also contained in Table 1) that do not agree with the experimental curve. The main problem with the *PMMA/POE* values is their higher error bars, associated with a sigmoidal curvature significantly more marked than that exhibited by the RPA results. This feature should be considered together with the significantly higher statistical noise associated with the charge equilibration method results. Finally, the previously

13obtained *PVME/PS* systems showed systematic differences between the results obtained with the two different methods to assign charges. In spite of these cautionary comments, Figure 5 shows a remarkably consistent pattern of quantitative agreement between experimental and simulation data of $\chi$, which supports the use of the present method.

In a previous work,[15] we have reported results obtained with a direct calculation of interaction energies and coordination numbers between monomer units in contact, proposed time ago.[16] This procedure includes a Monte Carlo generation of differently oriented monomer conformations. The results from these calculations were in general qualitative agreement with the experimental behavior of the blends studied in the present work, though the absolute values of $\chi$ were clearly overestimated in all cases. Moreover, an abrupt increase of $\chi$ with temperature was predicted for the *PMMA/POE* blends. Furthermore, the performance of this method was especially poor in the case of *PVME/PS* blends, for which high positive and temperature decreasing values of $\chi$ were obtained.

We also considered an alternative computationally simpler method, based in a basic theoretical treatment that has allow us to relate a particular form of binary interaction integrals with parameter $\chi$.[15] This method is also implemented by performing Monte Carlo orientational averages for pairs of our short chain conformations placed at different distances. The results are in general qualitative agreement with the experimental data for all systems, including the *PVME/PS* chains. Also, a fair quantitative agreement is found in most systems, though the method yields significantly higher values of $\chi$ for the *PMMA/PS* blends. However, it should be noted these satisfactory results have been obtained after some modifications of the forcefield parameters with respect to the default values for condensed phase used in the present MD simulations. These forcefield modifications were common for all the systems. Specifically, we chose a fixed shorter cut-off for both the Van der Waals and Coulombic interactions and we included a carefully fitted distance-dependent dielectric constant. Taking into account the shortcomings of these simplified Monte Carlo method, it can be argued that the present MD calculations provide a more general and reliable method to estimate $\chi$, even though they require a greater computational effort.

In our previous report of MD simulations for the *PVME/PS* blend,[8] we also discussed alternative calculations for $\chi$ obtained through the evaluation of a cohesive energy density or solubility parameter.[21] In comparison with the experimental data, these results also showed negative $\chi$ estimations, though with higher absolute values and a negative temperature variation. An important general advantage of obtaining $\chi$ from scattering functions is that this method involves calculations involving unit coordinates, without directly using energy values. Consequently, the influence of the forcefield details is minimized.

**Acknowledgment** This work has been partially supported by Grant CTQ2006-06446 from DGI-MEC Spain.

**Table 1.- Simulation values of $\chi/V_0$ for different blends, obtained with the two alternative methods to assign the partial charges, see text.**

| T(K) | Forcefield assigned | Charge equilibration method |
|---|---|---|
| *PMMA/PS* | | |
| 350 | 0.006±0.005 | 0.007±0.005 |
| 400 | 0.009±0.007 | 0.002±0.010 |
| 450 | 0.005±0.004 | 0.001±0.006 |
| *PS/PI[a]* | | |
| 350 | 0.007±0.010 | 0.002±0.011 |
| 400 | 0.002±0.006 | -0.001±0.007 |
| 450 | -0.002±0.010 | -0.004±0.011 |
| *PS/PI[b]* | | |
| 350 | 0.022±0.008 | 0.023±0.005 |
| 400 | 0.018±0.005 | 0.010±0.004 |
| 450 | 0.015±0.008 | 0.014±0.006 |
| *PMMA/POE* | | |
| 350 | -0.004±0.010 | -0.015±0.018 |
| 400 | -0.008±0.013 | 0.004±0.018 |
| 450 | 0.006±0.008 | -0.006±0.013 |

[a] All points considered. [b] Neglecting the last three points, see text.

**Figure captions**

**Figure 1**. $1/S(q)$ vs. $q^2$ from the MD simulation of a *PMMA/PS* mixture at 400 K. Partial charges are assigned by the forcefield. Symbols: simulation data; curves correspond to the RPA results from eq.(4), from bottom to top: dotted line, $\chi/V_0$ =0.05; solid line: $\chi/V_0$ =0; dashed line, $\chi/V_0$ =-0.05.

**Figure 2**. $1/S(q)$ vs. $q^2$ from the MD simulation of a *PI/PS* mixture at 400 K. Partial charges are assigned by the forcefield. Symbols: simulation data; curves correspond to the RPA results from eq.(4), from bottom to top: dotted line, $\chi/V_0$ =0.05; solid line: $\chi/V_0$ =0; dashed line, $\chi/V_0$ =-0.05.

**Figure 3**. $1/S(q)$ vs. $q^2$ from the MD simulation of a *PMMA/POE* mixture at 400 K. Partial charges are assigned by the forcefield. Symbols: simulation data; curves correspond to the RPA results from eq.(4), from bottom to top: dotted line, $\chi/V_0$ =0.05; solid line: $\chi/V_0$ =0; dashed line, $\chi/V_0$ =-0.05.

**Figure 4**. $1/S(q)$ vs. $q^2$ from the MD simulation of a *PMMA/POE* mixture at 400 K. Partial charges are obtained according to the charge equilibration method.[18] Symbols: simulation data; curves correspond to the RPA results from eq.(4), from bottom to top: dotted line, $\chi/V_0$ =0.05; solid line: $\chi/V_0$ =0; dashed line, $\chi/V_0$ =-0.05.

**Figure 5**. Parameter $\chi$ (referred to a volume of 100 Å$^3$) vs. temperature. Mean values of the simulation results, obtained from eq.(7) are denoted by symbols (with error bars). Squares: results obtained previously for the *PVME/PS* system.[8]; circles:

*PMMA/PS*; triangles: *PI/PS*, neglecting last thee points, see text; inverted triangles: *PMMA/POE*. Lines correspond to the reported fitted equation to neutron scattering experimental data.[3] Solid: *PVME/PS*; dashed: *PMMA/PS*; dotted: *PI/PS*; dashed-dotted: *PMMA/POE*.

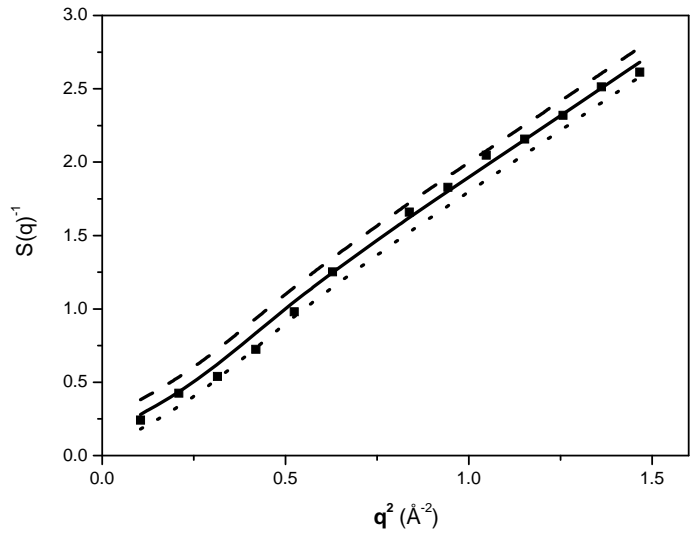

Figure 1

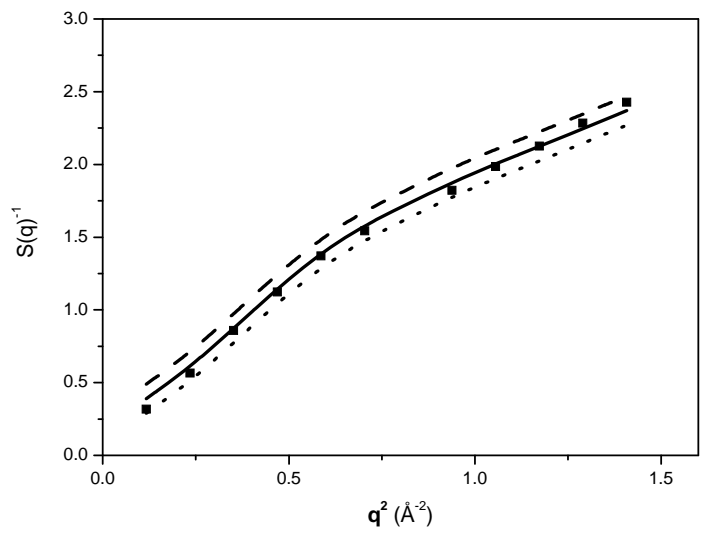

Figure 2

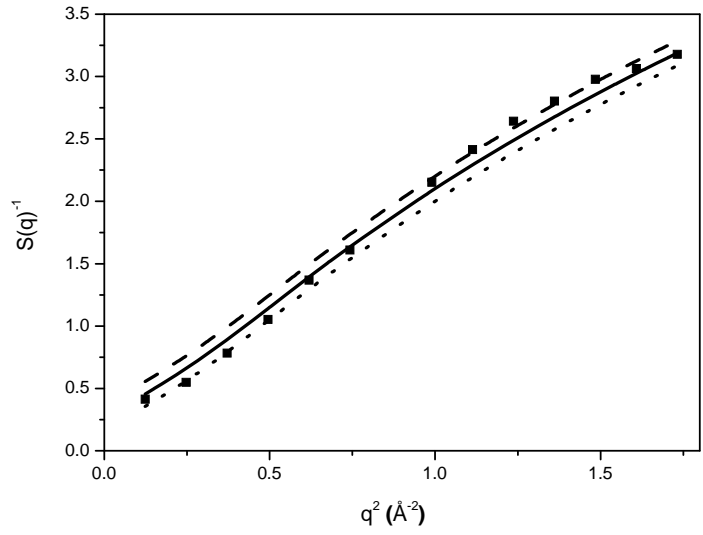

Figure 3

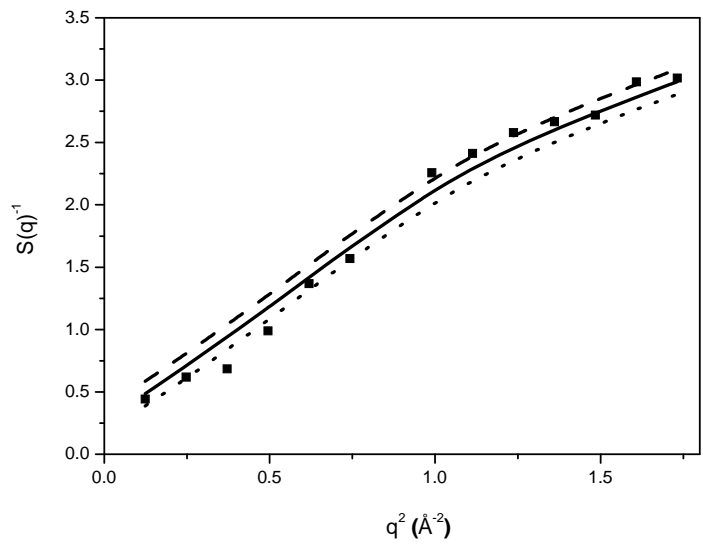

Figure 4